 \documentclass[letterpaper, 10 pt, conference]{ieeeconf}
\IEEEoverridecommandlockouts                              

\usepackage{cite}
\expandafter\let\csname proof\endcsname\relax
\expandafter\let\csname endproof\endcsname\relax
\usepackage{amsmath,amssymb,amsfonts}
\usepackage{graphicx}
\usepackage{textcomp}
\usepackage{xcolor}
\usepackage{amsthm}
\usepackage{hyperref}
\usepackage[ruled]{algorithm2e} 
\usepackage{algorithmic}
\usepackage{booktabs}
\usepackage{balance}
\def\BibTeX{{\rm B\kern-.05em{\sc i\kern-.025em b}\kern-.08em
    T\kern-.1667em\lower.7ex\hbox{E}\kern-.125emX}}
    \newtheorem{theorem}{Theorem}

\newtheorem{lemma}{Lemma}

\newtheorem*{remark}{Remark}
\newtheorem{definition}{Definition}
\newtheorem{proposition}{Proposition}
\newtheorem{assumption}{Assumption}
\usepackage{threeparttable}

\title{\LARGE \bf
Safety-Aware Optimal Control for Motion Planning with Low Computing Complexity
}
\author{Xuda Ding$^{1*}$, Han Wang$^{2*}$, Jianping He$^{1}$, Cailian Chen$^{1}$, Kostas Margellos$^{2}$, Antonis Papachristodoulou$^{2}$
\thanks{$^{*}$Co-primary authors}
\thanks{$^{1}$The authors are with the Department of Automation Shanghai Jiao Tong University, Shanghai, China.
E-mails: {\tt\small \{dingxuda, jphe,cailianchen\}@sjtu.edu.cn}}%
\thanks{$^{2}$The authors are with the Department of Engineering Science, University of Oxford, Oxford, United Kingdom. E-mails: {\tt\small \{han.wang, kostas.margellos, antonis\}@eng.ox.ac.uk}}
}

\begin{document}

\maketitle
\thispagestyle{empty}
\pagestyle{empty}

\begin{abstract}


The existence of multiple irregular obstacles in the environment introduces nonconvex constraints into the optimization for motion planning, which makes the optimal control problem hard to handle.
One efficient approach to address this issue is Successive Convex Approximation (SCA), where the nonconvex problem is convexified and solved successively. 
However, this approach still faces two main challenges:
\romannumeral1) infeasibility, caused by linearisation about infeasible reference points;
\romannumeral2) high computational complexity incurred by multiple constraints, 
when solving the optimal control problem with a long planning horizon and multiple obstacles.
To overcome these challanges, this paper proposes an energy efficient safety-aware control method for motion planning with low computing complexity and address these challenges.
Specifically, a control barrier function-based linear quadratic regulator is formulated for the motion planning to guarantee safety and energy efficiency.
Then, to avoid infeasibility, Backward Receding SCA (BRSCA) approach with a dynamic constraints-selection rule is proposed.
Dynamic programming with primal-dual iteration is designed to decrease computational complexity.
It is found that BRSCA is applicable to time-varying control limits. 
Numerical simulations and hardware experiments vevify the efficiency of BRSCA.
Simulations demonstrates that BRSCA has a higher probability of finding feasible solutions, reduces the computation time by about 17.4\% and the energy cost by about four times compared to other methods in the literature.


\end{abstract}
\section{Introduction}

Motion planning is critical for robotics since it provides fundamental movement guidance.
Many efforts are proposed to perform an efficient and collision-free planning.
The existing motion planning approach can be grouped into two categories \cite{schulman2014motion}, i.e., graph search approaches and trajectory optimization approaches. 
Graph search approaches are widely used in 2-D scenarios by discretizing the working space and efficiently finding collision-free trajectories \cite{duchovn2014path, yao2010path}. 
When applied to high-dimensional spaces, these approaches often encounter real-time problems since the search space is enormous.
To deal with real-time problems, Rapidly-Exploring Random Trees (RRT) \cite{rybus2020point,thakar2022manipulator} and Probabilistic Road Map (PRM) \cite{kavraki1996probabilistic} with pruning technique are proposed.
These methods are probabilistically complete, which means they may take a long time to achieve asymptotic optimality.
Optimization methodologies are proposed to deal with optimality, which also scales well with the state space dimension. 
However, the long time horizon and the nonconvex collision avoidance constraints make the optimization problem hard to find an optimal solution \cite{betts1998survey,bjelonic2022offline}.
Even a feasible solution is hard to be determined with a nonlinear optimization solver like Sequential Quadratic Programming (SQP) \cite{banerjee2020learning}. To overcome these problems, two different approaches have been proposed.


The first involves partitioning the safety region into a series of convex regions. 
It has been proposed to use polyhedra for the region generation, and then the trajectory optimization was formulated as a mixed-integer programming problem \cite{deits2015computing} \cite{deits2015efficient}. 
Polynomials and splines were also considered in more recent works \cite{tordesillas2019faster} \cite{tordesillas2021mader}. 
These approaches are highly efficient for finding a feasible trajectory without considering system dynamics and controller design, i.e., stability and optimal energy consumption.  


The second stream of methods gradually convexifies the nonconvex collision avoidance constraints via linearization. 
This type of approach was firstly proposed for the difference of convex programming problems \cite{le2018dc}. 
These approaches split the nonconvex function into the difference of two convex parts, then successfully convexify the constraints' nonconvex parts via linearisation about a reference point, 
thus termed Successive Convex Approximation (SCA). 
Variations of such approaches are proposed in the optimization community in the realm of the convex-concave procedure \cite{lipp2016variations}. 
This efficient transformation has been widely used in many recent optimization-based works, e.g., \cite{liu2018convex,mao2017successive,szmuk2020successive}. 
Although mathematically rigorous, the SCA still faces critical issues in real applications. 
The first one is that a feasible initial guess is required. 
Slack variables are used to relax this issue \cite{lipp2016variations}, but it still lacks a theoretical guarantee. 
The second issue is that the search space is prone to be empty due to the presence of a high number of constraints. 
An incremental SCA (iSCA) approach has been proposed, which incrementally includes the violated constraints into the optimization problem and guarantees a lower computation complexity \cite{chen2015decoupled}. 
However, in some multi-constraint scenarios, the trajectory calculated by iSCA is infeasible. The reason is that the convexification around infeasible reference points possibly renders the convex search spaces infeasible. 

To design stable and efficient control laws for safe motion planning, model predictive control and optimal control with collision avoidance constraints are widely investigated.
Moreover, the QP-based control barrier function (CBF) approach is further proposed to enforce safety \cite{thakar2022manipulator, singletary2022safety}.
Modern nonlinear solvers enable us to reformulate the problem into a generalized optimization problem with an equality constraint (system dynamics) and inequality constraints (collision avoidance and input limits). The computational complexity of such a problem grows quadratically with the length of the planning horizon and the number of obstacles \cite{augugliaro2012generation}. So even a linear quadratic regulator (LQR) model with multiple convexified constraints is challenging.
Many efforts have been made to solve this problem with multi-parametric optimization, which partitions the state space and substitutes multiple constrained LQR sub-problems for the original problem \cite{bemporad2002explicit,scokaert1998constrained,ferranti2016constrained}. 
Recently, \cite{chen2020optimal} proposed a density function-based approach that generates a control law for whole state space, whereas the approach requires solving complicated, ordinary differential equations in every iteration.

This paper proposes a novel primal-dual framework for solving the LQR with CBF constraints through backward recursion to obtain a safe and optimal motion planning in a long horizon. 
First, we formulate an LQR considering the system dynamic with zeroing barrier function to guarantee safety.
The obstacles are considered nonconvex constraints.
Then, all the nonconvex constraints are convexed with the proposed Backward Receding SCA (BRSCA) scheme to reduce the computing complexity.
The original problem is transformed into a convex Quadratically Constrained Quadratic Programming (QCQP). The proposed approach provides an explicit solution for the optimal cost-to-go and the control law.
Time-varying control limits are also considered in the controller synthesis.
Our work provides a solution for the sub-problem in \cite{stathopoulos2016solving}. This work is mainly related to the recent constrained differential dynamic programming approach \cite{aoyama2020constrained} \cite{plancher2017constrained}. Unlike lifting the constraints into the cost with barrier functions, we model the constraints in a hard manner. In addition, our method is especially suitable for linear systems, as the cost-to-go can be accurately modelled.
The contributions can be summarised as follows:
\begin{itemize}
    \item \textit{Feasibility: }
    We propose a novel BRSCA approach to deal with infeasible reference points and provide acceleration mechanisms.
    BRSCA guarantees the feasibility when solving a CBF-based LQR in a long time horizon.
    \item \textit{Efficiency: } 
    We propose a primal-dual framework for solving convex-constrained LQR, with closed-form solutions for the optimal cost-to-go and the safety-aware control law.
    
    \item \textit{Pracitcal implementation: } We demonstrate higher computational efficiency and success rate of our method against five solvers and five planning methods through a detailed numerical study and real-world implementation.
\end{itemize}

The rest of this paper is organized as follows.
Section \ref{sec:formulation} formulates the optimal control problem with constraints and provides the definition of a semi-convex function. 
BRSCA for convexifying the original problem is presented in Section \ref{sec:brsca}.
The explicit solution of the optimal control problem is shown in Section \ref{sec:primal-dual}.
Section \ref{sec:simulation} presents the numerical simulations and hardware implementations. Section \ref{sec:conclusion} concludes the paper and provides some directions for future work.

\section{Problem Formulation}\label{sec:formulation}
Consider a robot with a configuration $\boldsymbol{x}_t \subset \mathbb{R}^n$,
 where $\mathbb{R}^n$ is the $n$-dimensional Euclidean space.
The robot dynamics is modelled as the following
\begin{equation}\label{model}
    \boldsymbol{x}_{t+1}=\boldsymbol{A}\boldsymbol{x}_t+\boldsymbol{B}\boldsymbol{u}_t,
\end{equation}
where $\boldsymbol{u}_t\in \mathcal{U}_t \subset \mathbb{R}^m$ denotes the control input, $\mathcal{U}_t$ is the time-varying allowable control set. $\boldsymbol{A}\in \mathbb{R}^{n\times n}$ is the state transition matrix, and $\boldsymbol{B} \in \mathbb{R}^{n \times m}$ is the input matrix. 
Assume that $\boldsymbol{A}$ and $\boldsymbol{B}$ are known, and $(\boldsymbol{A}, \boldsymbol{B})$ is stabilizable.

The objective is to synthesize an optimal control law along with an optimal trajectory for a single robot. 
To avoid collision, all states $\{\boldsymbol{x}_0,\ldots,\boldsymbol{x}_t,\ldots\}$ should be within a safety set $\mathcal{X}$ defined by the zeroing barrier function proposed in \cite{prajna2004safety}. Suppose there exist a series of functions $h_i:\mathbb{R}^n\to \mathbb{R}$ to define the safety set $\mathcal{X}$,
\begin{equation}\label{barrier}
    \begin{split}
        \mathcal{X}&=\cap_{i \in {\cal I}} {\{\boldsymbol{x}\in\mathbb{R}^n|h_i(\boldsymbol{x})\ge 0\}}, \\
     \partial\mathcal{X}&=\cup_{i \in {\cal I}} {\{\boldsymbol{x}\in\mathbb{R}^n|h_i(\boldsymbol{x})= 0\}},\\
        \text{Int}(\mathcal{X})&=\cap_{i \in {\cal I}} {\{\boldsymbol{x}\in\mathbb{R}^n|h_i(\boldsymbol{x})> 0\}},\\
        \bar{\mathcal{X}}&=\cup_{i \in {\cal I}} {\{\boldsymbol{x}\in\mathbb{R}^n|h_i(\boldsymbol{x})< 0\}},
    \end{split}
\end{equation}
where $\partial \mathcal{X}$ represents the boundary of safety set, Int($\mathcal{X}$) and $\bar{\mathcal{X}}$ are the interior and complementary set of $\mathcal{X}$, respectively. A sequence of functions $h_1(\boldsymbol{x}),\ldots, h_\mathcal{I}(\boldsymbol{x})$ can be used to establish the safety set and obstacle descriptions where every obstacle is specified by a unique index in $\mathcal{I}$. With a slight abuse of notation, let the subscript $\mathcal{I}$ represent the last index. Assume all these sets are smooth, compact, and \textit{semi-convex}.

\begin{definition}[\textit{Semi-convex}]\label{def:semi-convex}
	Let $\mathcal{H}_i(\boldsymbol{x},\boldsymbol{x}_0)$ be a quadratic function with a positive semidefinite matrix ${\boldsymbol{H}}_i\in\mathbb{R}^{n\times n}$,
	\begin{equation}
		\mathcal{H}_i(\boldsymbol{x},\boldsymbol{x}_0):= \frac{1}{2}(\boldsymbol{x}-\boldsymbol{x}_0)^\top {\boldsymbol{H}}_i(\boldsymbol{x}-\boldsymbol{x}_0)
	\end{equation}
	Then, a function $h_i(\boldsymbol{x})$ is said to be semi-convex if the following equality holds.
	\begin{equation}\label{semiconvex}
		 h_i(\boldsymbol{x})  \buildrel \Delta \over = \tilde h_i(\boldsymbol{x}) - \mathcal{H}_i(\boldsymbol{x},\boldsymbol{x}_0),
	\end{equation}
	\end{definition}
\begin{assumption}\label{assum:semi-convex}
All functions $\{-h_1(\boldsymbol{x}),\ldots,-h_\mathcal{I}(\boldsymbol{x})\}$ in \eqref{barrier} are \textit{semi-convex}.
\end{assumption}
We note here we assume $-h_i(\boldsymbol{x})$ is \textit{semi-convex} but not $h_i(\boldsymbol{x})$, since the collision-free constraints can be reformulated as $-h_i(\boldsymbol{x})\le 0$.

\begin{remark}
Semi-convex obstacles are quite common in practice. Roughly speaking, for every feasible state, if there exists a convex quadratic closure that covers it without intersecting with the obstacle, we say the function corresponding to the obstacle is \textit{semi-convex}.
Nonconvex obstacles can often be approximated/decomposed as the union of convex quadratic closure\cite{zhang2020optimization}.
\end{remark}


Unlike QP-based control barrier function approaches, which only depend on the current state, we focus here on transforming the problem into an optimal control problem with collision avoidance constraints at every time instant.
\vspace{-20pt}
	\begin{subequations}\label{SCLQR}
		\begin{align}
\mathop {\min }\limits_{\boldsymbol{u}\in \mathbb{R}^n}~~~&J(\boldsymbol{u})={\boldsymbol{x}_T}^\top{\boldsymbol{P}}{\boldsymbol{x}_T} + \sum\limits_{t = 0}^{T - 1} {{\boldsymbol{x}_t}^\top {\boldsymbol{Q}}{\boldsymbol{x}_t} + {\boldsymbol{u}_t}^\top {\boldsymbol{R}}{\boldsymbol{u}_t}}, \label{SCLQR-a}\\
s.t.~~~&{\boldsymbol{x}_{t+1}} = \boldsymbol{A}{\boldsymbol{\boldsymbol{x}}_t} + \boldsymbol{B}{\boldsymbol{u}_t},t = 0, \ldots ,T - 1,\label{SCLQ-b}\\
&h_i({\boldsymbol{x}_t}) \ge 0,t = 1, \ldots ,T-1,i\in\mathcal{I},\label{SCLQR-c}\\
&\mathcal{G}_t (\boldsymbol{u}_t)= {\boldsymbol{G}_t}\boldsymbol{u}_t+{\boldsymbol{e}_t}\le 0,t=0,\ldots,T-1,\label{SCLQR-d}
		\end{align}
	\end{subequations}
where $J$ denotes the energy cost, $\boldsymbol{P}\succeq 0$ denotes the terminal cost function, ${\boldsymbol{Q}}\succeq 0$ and ${\boldsymbol{R}}\succ 0$ denote the state and input weights, respectively.
$(\boldsymbol{A},\sqrt{{\boldsymbol{Q}}})$ is detectable and
$\mathcal{G}_t (\boldsymbol{u}_t) \le 0$ formulates the time-varying control admissible set $\mathcal{U}_t$ (where ${\boldsymbol{G}_t} \in \mathbb{R}^{s\times n}, {\boldsymbol{e}_t}\in \mathbb{R}^s$). 
Note that the terminal term ${\boldsymbol{x}_T}^\top \boldsymbol{P}\boldsymbol{x}_T$ is used to reach and stabilize around the equilibrium. The length of the horizon $T$ depicts the trade-off between computational complexity and conservatism. 



Moreover, the obstacles are detected by a centralized perception unit in practice.
The shape functions of the obstacles can be regressed by sampling the obstacles' shapes.
Such procedures are broadly used in graph search methods.
When only onboard sensors are used for detecting the obstacles, the constraints for \eqref{SCLQR} are not complete.
Thus, the optimality of motion planning is not guaranteed.
As mentioned in our motivation, our goal is to find a feasible solution for the constrained problem \eqref{SCLQR} in time.

\section{Backward Receding SCA}\label{sec:brsca}
Since the unnecessarily convex constraint set $h_i(\boldsymbol{x}_t) \ge 0$ makes it hard to leverage dynamic programming to construct a backward iterative law, the feasibility is hard to achieve.
This subsection introduces the BRSCA approach to deal with feasibility problem.
Unlike iSCA, BRSCA checks the feasibility of all constraints and includes only the violated ones in the problem to guarantee the feasibility of the optimization problem with multi-constraint.

\subsection{SCA}

SCA was proposed for convexifing the nonconvex collision avoidance constraint $h_i(\boldsymbol{x}_0)\ge 0$ about state $\boldsymbol{x}_0$. The principle of SCA is to split the nonconvex function into the sum of a convex and a concave function and successively linearise the concave one about the current state. We can obtain the summation of the convex and concave parts by exploiting semi-convexity. Consider one nonconvex function $h_i(\boldsymbol{x})$ with reference point $\boldsymbol{x}_0$. Then the explicit expression of one convexified candidate $\hat h_i(\boldsymbol{x})$ takes the form
\begin{equation}\label{linearization}
\begin{split}
 {\hat h_i}(\boldsymbol{x}) = \underbrace { - {h_i}({\boldsymbol{x}_0}) - \nabla {h_i}{{({\boldsymbol{x}_0})}^\top}(\boldsymbol{x} - {\boldsymbol{x}_0})}_{{\text{linearized concave part}}} + \underbrace {\mathcal{H}_i(\boldsymbol{x},\boldsymbol{x}_0)}_{{\text{convex part}}}.
\end{split}
\end{equation}
\begin{lemma}
Assume that $\hat h_i(\boldsymbol{x})$ is convex, then the safe set defined by $\{\boldsymbol{x}|-\hat h_i(\boldsymbol{x})>0\}$ is a subset of that defined by $\{\boldsymbol{x}|h_i(\boldsymbol{x})>0\}$, and $-\hat h_i(\boldsymbol{x})\le h_i(\boldsymbol{x}), \forall \boldsymbol{x} \in \mathbb{R}^n$.
\end{lemma}
\begin{proof}
By Definition \ref{def:semi-convex}, $h_i(\boldsymbol{x})+\mathcal{H}_i(\boldsymbol{x},\boldsymbol{x}_0)$ is convex since $h_i(\boldsymbol{x})$ is assumed to be \textit{semi-convex}. Hence we can substitute \begin{equation}
    -h_i(\boldsymbol{x})-\mathcal{H}_i(\boldsymbol{x},\boldsymbol{x}_0)+\mathcal{H}_i(\boldsymbol{x},\boldsymbol{x}_0)<0
\end{equation}
for the safe constraint $-h_i(\boldsymbol{x})<0$. 
Note that $-h_i(\boldsymbol{x})-\mathcal{H}_i(\boldsymbol{x},\boldsymbol{x}_0)$ and $\mathcal{H}_i(\boldsymbol{x},\boldsymbol{x}_0)$ are concave and convex, respectively. By the definition of convexity, we obtain:
\begin{equation}
    \begin{split}
        h_i(\boldsymbol{x})-h_i(\boldsymbol{x}_0)\ge
        -\mathcal{H}_i(\boldsymbol{x},\boldsymbol{x}_0)+\nabla h_i(\boldsymbol{x}_0)(\boldsymbol{x}-\boldsymbol{x}_0),
    \end{split}
\end{equation}
which implies that 
\begin{equation}
    \begin{split}
        h_i(\boldsymbol{x})\ge
        h_i(\boldsymbol{x}_0)-\mathcal{H}_i(\boldsymbol{x},\boldsymbol{x}_0)+\nabla h_i(\boldsymbol{x}_0)(\boldsymbol{x}-\boldsymbol{x}_0)=-\hat h_i(\boldsymbol{x}),
    \end{split}
\end{equation}
i.e., $-\hat h_i(\boldsymbol{x})\le h_i(\boldsymbol{x})$ holds. Besides, since $\mathcal{H}_i(\boldsymbol{x},\boldsymbol{x}_0)$ is convex and $-h_i(\boldsymbol{x})-\nabla h_i(\boldsymbol{x}_0)^\top(\boldsymbol{x}-\boldsymbol{x}_0)$ is affine, $\hat h_i(\boldsymbol{x})$ is convex.
\end{proof}

We obtain the convexified expressions for each constraint through convexification \eqref{linearization}. However, when the amount of obstacles is large, the search space can be highly limited, and even the feasibility of the problem is not guaranteed. 

\subsection{BRSCA}
	\begin{algorithm}[t]
		\LinesNumbered
		\caption{BRSCA algorithm}\label{al:CPSCAA}
		\KwIn{length of planning horizon $T$, safe set $\mathcal{X}$, initial start point $\boldsymbol{x}_0$, end point $\boldsymbol{x}_T$,  trajectory $\{\boldsymbol{x}_0^0,\ldots,\boldsymbol{x}_T^0\}$ solved from \eqref{SCLQR} without collision free constraints
		}
		\KwOut{optimal control law $\boldsymbol{u}^*$, optimal cost value $J^*(\boldsymbol{u}^*)$.
		\vspace{3pt}	
		}
		\While{$\exists i,t,~h_i(\boldsymbol{x}_t^k)\le 0$}
		{
		include the violated constraints $h_i(\boldsymbol{x}_t^k)\le0$.\\
		    \While{$J(\boldsymbol{u}^{k+1})<J(\boldsymbol{u}^k)$}
		    {
		    \For{all the non-violated constraints $h_j(\boldsymbol{x}_t^k)>0$}
		    {convexify \eqref{SCLQR-c} according to \eqref{linearization}
		    }
		    \For{all the violated constraints $h_j(\boldsymbol{x}_t^k)>0$}
		    {
		    convexify \eqref{SCLQR-c} according to \eqref{linearization} with the closest backward feasible state $\boldsymbol{x}_{\hat t}^k$ as the new reference point $\boldsymbol{x}_0$, i.e. $\hat t=\arg \min_{\hat t}({h_i(\boldsymbol{x}_{\hat t}^k)>0})$, $\hat t<t$
		    }
		    solve $\boldsymbol{u}^*$, $J^*(\boldsymbol{u}^*)$, and the new $\{\boldsymbol{x}_0^{k+1},\ldots,\boldsymbol{x}_T^{k+1}\}$
		    }
		}
	\end{algorithm}
BRSCA is introduced based on a backward receding scheme, shown in Algorithm \ref{al:CPSCAA}, to enlarge the feasible search space, as well as increase the success rate. 
\eqref{SCLQR} is with $\mathcal{O}(T\hat{\mathcal{I}})$ inequality constraints in BRSCA, where $T$ is the number of discretization steps and $\hat{\mathcal{I}}$ is the number of the constraints included in the algorithm.
It should be noticed that $\hat{\mathcal{I}}$ is less than the number of the obstacle $\mathcal{I}$.
The runtime of the BRSCA is $\mathcal{O}(T^2\hat{\mathcal{I}}^2)$ \cite{chen2015decoupled}.

\begin{remark}
Step 3 of Algorithm \ref{al:CPSCAA} is similar to iSCA in that the constraints are included dynamically. Only violated collision avoidance constraints are included in the constrained LQR problem, while we only consider the included constraints in the problem.
Steps 4 - 8 of Algorithm \ref{al:CPSCAA} show the BRSCA. Here we introduce a novel approach of using the convexified search space about the closest backward feasible state, instead of that about an infeasible state. This efficient rule eliminates the feasible initial guess requirement of the iSCA, and reduces the number of the iteration and computing complexity. In step 10 we solve the constrained LQR problem for optimal control law $\boldsymbol{u}^*=\{\boldsymbol{u}_0^*,\ldots,\boldsymbol{u}_{T-1}^*\}$ and obtain the corresponding trajectory, and the optimal cost value $J^*(\boldsymbol{u}^*)$.  
\end{remark}

Figure \ref{fig:convexsearchspace} shows the convex search space about the violated reference point $\hat{\boldsymbol{x}}_1$ for a semi-convex obstacle. After including the constraints about $\hat{\boldsymbol{x}}_1$ and convexifying those via the search space of the closest feasible backward point $\boldsymbol{x}_0$, we re-solve the problem and acquire the new trajectory (denoted by the red line), which is feasible at $\boldsymbol{x}_1$, while the previous trajectory (denoted by the dotted line) is infeasible at $\hat{\boldsymbol{x}}_1$. 

In the sequel, we show how to solve the constrained LQR problem in Step 10 efficiently.

\section{Primal-Dual Controller Synthesis}\label{sec:primal-dual}
This section shows the explicit solution of the constrained LQR problem, which ensures the computing efficiency.
After convexifying all the nonconvex constraint functions via \eqref{linearization}, we reformulate \eqref{SCLQR} into a convex QCQP by convexifying \eqref{SCLQR-c} with
\begin{equation}\label{RLQR}
\begin{split}
        f_i(\boldsymbol{x}_t) &= {\boldsymbol{x}_t}^\top {\boldsymbol{H}}_i{\boldsymbol{x}_t} + {{\boldsymbol{c}}_{i|t}}^\top {\boldsymbol{x}_t} + {d_{i|t}}\le 0,t \in \mathcal{T},i \in \mathcal{I}_t,\\
    {\boldsymbol{c}}_{i|t}^\top&=-\nabla h_i(\boldsymbol{x}_t^k)^\top-(\boldsymbol{x}_t^k)^\top {\boldsymbol{H}}_i,\\
    d_{i|t}&=-h_i(\boldsymbol{x}_t^k)+\nabla h_i(\boldsymbol{x}_t^k)^\top \boldsymbol{x}_t^k+\frac{1}{2}(\boldsymbol{x}_t^k)^\top {\boldsymbol{H}}_i\boldsymbol{x}_t^k,
    \end{split}
\end{equation}
where $\mathcal{T},\mathcal{I}_t$ represent the time and index set of the included constraints, respectively. $\boldsymbol{x}^k$ represents the $k$-th solution in Step 11 of the Algorithm \ref{al:CPSCAA}. For brevity, we omit the index $k$ in \eqref{RLQR}. In the sequel, $\boldsymbol{x}^k$ is used as the $k$-th primal variable in the primal-dual iteration.


\subsection{Lagrangian Duality Formulation}
To solve \eqref{SCLQR} with the convexified constraints \eqref{RLQR}, consider the Lagrangian $L(\boldsymbol{u},\lambda,{\boldsymbol{\mu}})$:
\begin{equation}\label{Lagrangian}
L(\boldsymbol{u},\lambda,{\boldsymbol{\mu}} ) = J(\boldsymbol{u}) + \sum\limits_{t = 1}^{T-1} {\sum\limits_{i\in\mathcal{I}_t} {{\lambda _{i|t}}f_i(\boldsymbol{x}_t)}}+\sum\limits_{t = 1}^{T - 1} {{{{\boldsymbol{\mu}} }_t}^\top \mathcal{G}_t(\boldsymbol{u}_t)},
\end{equation}
where $\lambda_{t|i}\in\mathbb{R}_+$ and ${\boldsymbol{\mu}}_t\in\mathbb{R}_+^s$ are the dual variables. $\boldsymbol{u}$, $\lambda$, and ${\boldsymbol{\mu}}$ are vectors consisting of $\boldsymbol{u}_t$, $\lambda_{i|t}$, and ${\boldsymbol{\mu}}_t$, respectively.
Accordingly, the dual function $D(\lambda,{\boldsymbol{\mu}})$ is defined as
\begin{equation}\label{dual}
   D(\lambda,{\boldsymbol{\mu}} ) = \mathop {\inf }\limits_{\boldsymbol{u}} L(\boldsymbol{u},\lambda,{\boldsymbol{\mu}} ).
\end{equation}

Then, from the duality theory \cite{boyd2004convex}, we have that 
\begin{equation}
    \mathop {\sup }\limits_{\lambda\ge 0,{\boldsymbol{\mu}} \ge0}  D(\lambda,{\boldsymbol{\mu}} ) \le \mathop {\inf }\limits_{\boldsymbol{u}} J(\boldsymbol{u}).
\end{equation}

We now present the optimality condition 
derived from the KKT conditions and strong convexity propositions.
\begin{proposition}\label{pro:KKT}
Suppose Slater's condition holds, i.e., there exists a series of inputs $\tilde u$ such that $\boldsymbol{x}_t^\top {\boldsymbol{H}}_i\boldsymbol{x}_t+{\boldsymbol{c}}_{i|t}^\top \boldsymbol{x}_t+d_{i|t}<0$ and ${\boldsymbol{G}_t}\tilde {\boldsymbol{u}}_t+{\boldsymbol{e}_t}< 0$. Then there exists an optimal control law $\boldsymbol{u}_t^*(\boldsymbol{x}_t,\lambda^*,{\boldsymbol{\mu}}^*)$ associated with dual variables $\lambda^*,{\boldsymbol{\mu}}^*$, which are defined as the maximum of Lagrangian $D(\lambda,{\boldsymbol{\mu}})$:
\begin{equation}
    {\lambda ^*,{\boldsymbol{\mu}}^*} = \mathop {\arg \max }\limits_{\lambda  \ge 0,{\boldsymbol{\mu}}\ge0} D(\lambda,{\boldsymbol{\mu}} ).
\end{equation}
	\begin{figure}[t]
    \centering
    \includegraphics[scale=0.8]{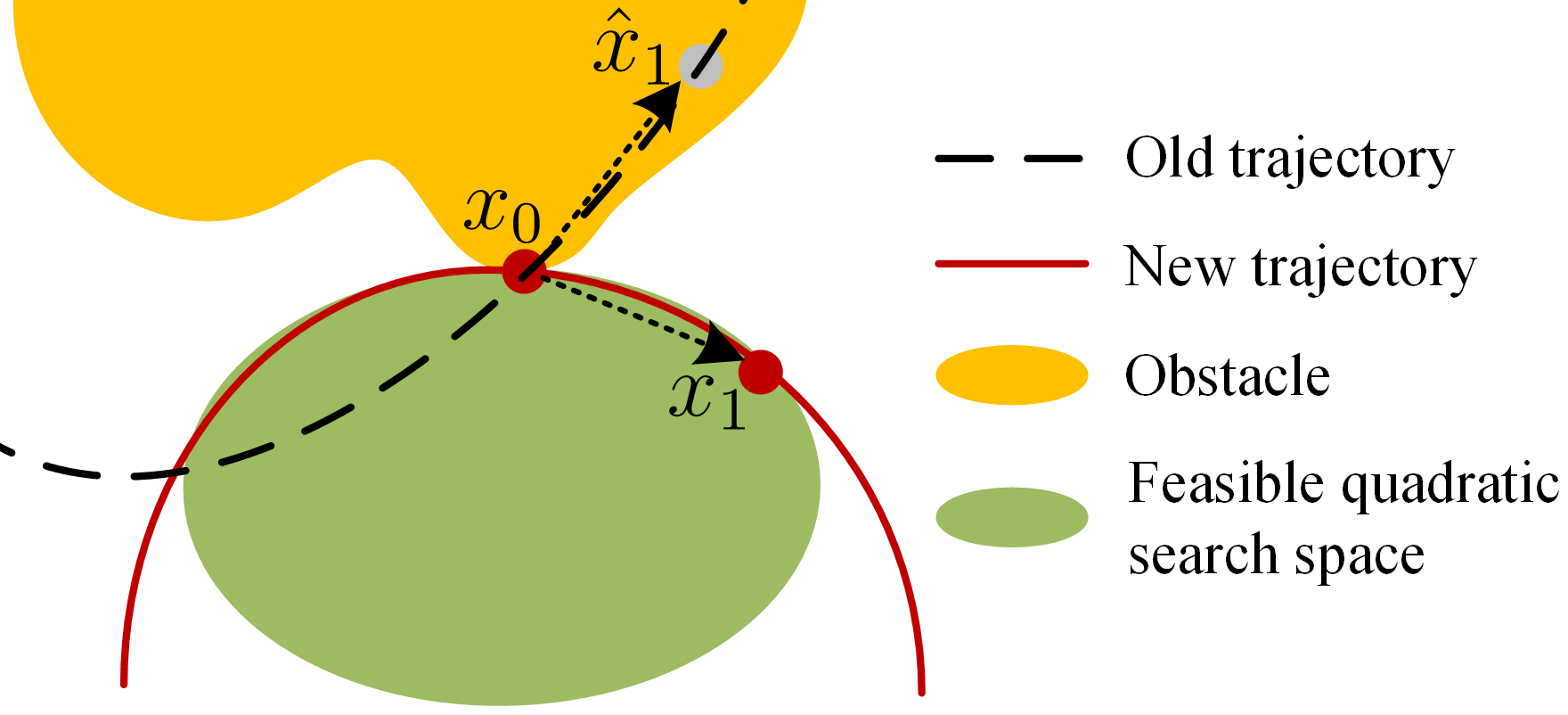}
    \caption{A demonstration of forming the convex search space and the new trajectory}
    \label{fig:convexsearchspace}
    \vspace{-10pt}
\end{figure}
Moreover, the following conditions hold:
\begin{itemize} 
    \item [1)] Duality gap is zero: 
    
    $L(\boldsymbol{u}^*,\lambda^*,{\boldsymbol{\mu}}^*)=\min_{\boldsymbol{u}}L(\boldsymbol{u},\lambda^*,{\boldsymbol{\mu}}^*)=D(\lambda^*,{\boldsymbol{\mu}}^*)$;
    \item [2)] Inequality constraints \eqref{SCLQR-d} and \eqref{RLQR} hold;
    \item [3)] Complementary slackness holds:
    
    $\lambda_{i|t}^*f_i(\boldsymbol{x}^*_t)=0$, $\forall t\in \mathcal{T},i\in \mathcal{I}_t$;
    
    ${\boldsymbol{\mu}}_t^*{\boldsymbol{G}_t}(\boldsymbol{u}_t^*)=0$, $t=0,\ldots,T-1$.
\end{itemize}
\end{proposition}

\begin{proof}
1) follows from the standard duality theorem, while 2) and 3) come from KKT conditions.
\end{proof}

\subsection{Primal-Dual Approach}\label{sub-PDA}
A primal-dual approach is used to solve the constrained LQR problem. The dual function $D(\lambda,{\boldsymbol{\mu}})$ is concave in $\lambda$ and ${\boldsymbol{\mu}}$. The gradient expressions $\nabla D(\lambda_{i|t})$ and $\nabla D({\boldsymbol{\mu}}_t)$ are:
\begin{equation}\label{gradient}
\begin{split}
       &\nabla D({\lambda _{i|t}}) = {\boldsymbol{x}_t}^\top{{\boldsymbol{H}}_i}{\boldsymbol{x}_t} + {{\boldsymbol{c}}_{i|t}}^\top{\boldsymbol{x}_t} + {d_{i|t}},\\
       &\nabla D({{\boldsymbol{\mu}}_t}) = {\boldsymbol{G}_t}\boldsymbol{u}_t+{\boldsymbol{e}_t}.
\end{split}
\end{equation}

Algorithm \ref{al:primal-dual} shows the primal-dual approach for solving the constrained LQR. The step sizes $\alpha^k$ fulfill that \romannumeral1) $\sum\limits_{k \to \infty } {{\alpha ^k} \to \infty } $; \romannumeral2) $\sum\limits_{k \to \infty } {{{({\alpha ^k})}^2} < \infty } $. Then through the convergence results of the dual ascent method, $[\boldsymbol{u}^k,\lambda^k,{\boldsymbol{\mu}}^k]$ converge to the saddle point of $L(\boldsymbol{u},\lambda,{\boldsymbol{\mu}})$ with sublinear convergence rate $\mathcal{O}(k)$ \cite{boyd2004convex,hamedani2021primal}. 

The quadratically convexified constraints enable us to solve the optimization sub-problem in Step 2. An explicit minimum can be derived through dynamic programming with every specific dual variable since there are no additional constraints other than system dynamics constraints in the sub-optimization problem.

\subsection{Optimal Safety-Critical Control Laws}
The solution to the optimization sub-problem in step 2 of Algorithm \ref{al:primal-dual} is shown in this subsection. The analysis uses the Hamilton-Jacobi-Bellman (HJB) equation and the Pontryagin Minimum Principle. We first define the auxiliary quadratic cost matrix
\begin{equation}
 {{\boldsymbol{Q}}_{\lambda|t} } \buildrel \Delta \over = {\boldsymbol{Q}} + \sum\limits_{i \in \mathcal{I}_t} { {{\lambda _{i|t}}{{\boldsymbol{H}}_i}} }. 
\end{equation}
where ${\boldsymbol{Q}}_{\lambda|t} \succeq 0$ since ${\boldsymbol{Q}}\succeq 0$ and ${\boldsymbol{H}}_i \succeq 0$.

For simplicity of notation in the sequel, the following substitutions are used
\begin{equation*}
    {\boldsymbol{\Lambda}}_t=\left[ {\begin{array}{*{20}{c}}
{{\lambda _{1|t}}}&{}& \cdots &{{\lambda _{\mathcal{I}_t|t}}}&{}\\
{}&{{\lambda _{1|t}}}& \cdots &{}&{{\lambda _{\mathcal{I}_t|t}}}
\end{array}} \right]^\top,
\end{equation*}
\begin{equation*}
    {\boldsymbol{C}}_t=[{\boldsymbol{c}}_{1|t},\ldots,{\boldsymbol{c}}_{{\mathcal{I}_t}|t}]^\top, {\boldsymbol{d}}_t=[d_{1|t},\ldots,d_{{\mathcal{I}_t}|t}]^\top,
\end{equation*}
and ${\boldsymbol{\lambda}}_t=[\lambda_{1|t},\ldots,\lambda_{\mathcal{I}_t|t}]^\top$.
$J(\boldsymbol{u})$ in \eqref{SCLQR-a} is replaced by $J(\boldsymbol{u},{\boldsymbol{\lambda}},{\boldsymbol{\mu}}) $ with \eqref{RLQR}:
\begin{equation}
     J(\boldsymbol{u},{\boldsymbol{\lambda}},{\boldsymbol{\mu}}) = g_T({\boldsymbol{x}},{\boldsymbol{\lambda}}) + \sum\limits_{t = 0}^{T - 1} {{g_t}({\boldsymbol{x}},{\boldsymbol{u}},{\boldsymbol{\lambda}},{\boldsymbol{\mu}})},
\end{equation}
where the terminal term $g_T$ and interval term $g_t$ are defined from \eqref{Lagrangian}:
\begin{equation}\label{costtogo}
    \begin{split}
g_T({\boldsymbol{x}}) &= {\boldsymbol{x}_T}^\top \boldsymbol{P}{\boldsymbol{x}_T},\\
g_t(\boldsymbol{x},\boldsymbol{u},{\boldsymbol{\lambda}},{\boldsymbol{\mu}})&={\boldsymbol{x}_t}^\top {\boldsymbol{Q}}_{\lambda|t} {\boldsymbol{x}_t}^\top+{\boldsymbol{C}}_t^\top{ {\boldsymbol{\Lambda}}_t}\boldsymbol{x}_t+{{\boldsymbol{\lambda}}_t}^\top {\boldsymbol{d}}_t\\
+&{\boldsymbol{\mu}}_t^\top({\boldsymbol{G}_t}\boldsymbol{u}_t+{\boldsymbol{e}_t})+{\boldsymbol{u}_t}^\top {\boldsymbol{R}}{\boldsymbol{u}_t}.
    \end{split}
\end{equation}

The quadratically convexified constraints enable us to exploit closed-form expressions for both minimum and optimal control law. Let ${V_t}^*(\boldsymbol{x},{\boldsymbol{\lambda}},{\boldsymbol{\mu}})$ denote the optimal cost-to-go with dual variables ${\boldsymbol{\lambda}},{\boldsymbol{\mu}}$ at time $t$:
\begin{equation}\label{value}
    {V_t}^*(\boldsymbol{x},{\boldsymbol{\lambda}},{\boldsymbol{\mu}} ) \buildrel \Delta \over = \mathop {\min }\limits_{\boldsymbol{u}} g_T({\boldsymbol{x}}) + \sum\limits_{k = t}^{T - 1} {{g_k}({\boldsymbol{x}},{\boldsymbol{u}},{\boldsymbol{\lambda}},{\boldsymbol{\mu}})}.
\end{equation}

We note here that the control admissible set $\mathcal{U}_t$ and the state admissible set $\mathcal{X}$ are not included here since these constraints have been lifted into the objective function. The minimum is always attained since $J(\boldsymbol{u},{\boldsymbol{\lambda}},{\boldsymbol{\mu}})$ is convex over $\boldsymbol{u}$.
An explicit expression of cost-to-go function ${V_t}^*(\boldsymbol{x},{\boldsymbol{\lambda}},{\boldsymbol{\mu}})$, and optimal control law ${\boldsymbol{u}_t}^*(\boldsymbol{x},{\boldsymbol{\lambda}},{\boldsymbol{\mu}})$ via dynamic programming with fixed dual variable ${\boldsymbol{\lambda}}$ and ${\boldsymbol{\mu}}$ is given in Theorem \ref{theorem1}. 
\begin{theorem}\label{theorem1}
With fixed dual variables ${\boldsymbol{\lambda}},{\boldsymbol{\mu}}$, for $t\le T-1$ the closed form expression of the optimal cost-to-go function ${V_t}^*(\boldsymbol{x},{\boldsymbol{\lambda}},{\boldsymbol{\mu}})$ is expressed as:
\begin{equation}
{V_t}^*({\boldsymbol{x}},{\boldsymbol{\lambda}},{\boldsymbol{\mu}}) = {\boldsymbol{x}_t}^\top{{\boldsymbol{F}}_t}{\boldsymbol{x}_t} + {{\boldsymbol{S}}_t}^\top{\boldsymbol{x}_t} + {r_t}.
\end{equation}

Moreover, the optimal control law associated with the dual variables ${\boldsymbol{\lambda}},{\boldsymbol{\mu}}$ is given by:
\begin{equation}\label{law}
\boldsymbol{u}_t^*(\boldsymbol{x},{\boldsymbol{\lambda}},{\boldsymbol{\mu}})=-{\boldsymbol{K}}_t^\top\boldsymbol{x}+l_t.
\end{equation}

The discrete-time backward recursions through dynamic programming are given by:
\begin{equation}\label{recursion}
    \begin{split}
        {{\boldsymbol{F}}_{t-1}} &=  - {{\boldsymbol{A}}^\top}{{\boldsymbol{F}}_{t }}{\boldsymbol{B}}{{\boldsymbol{M}}_{t-1}^{-1}}{{\boldsymbol{B}}^\top}{{\boldsymbol{F}}_{t}}{\boldsymbol{A}} + {{\boldsymbol{Q}}_{{\lambda}|t} } + {{\boldsymbol{A}}^\top}{{\boldsymbol{F}}_{t}}{\boldsymbol{A}},\\
        {\boldsymbol{S}}_{t - 1}^\top &= {\boldsymbol{C}}_{t-1}^\top{{\boldsymbol{\Lambda}} _{t - 1}} + {\boldsymbol{S}}_t^\top {\boldsymbol{A}}\\
                    & - ({\boldsymbol{S}}_t^\top {\boldsymbol{B}}+{\boldsymbol{\mu}}_{t-1}G_{t- 1}){{\boldsymbol{M}}^{ - 1}}{{\boldsymbol{B}}^\top }{{\boldsymbol{F}}_t}^\top {\boldsymbol{A}},\\
        {r_{t - 1}} &= {{\boldsymbol{\lambda}} _{t - 1}}^\top {\boldsymbol{d}}_{t-1} -{\boldsymbol{\mu}}_{t-1}{\boldsymbol{e}_{t-1}}+ {r_t},\\
        {\boldsymbol{M}}_{t-1}&={{\boldsymbol{B}}^\top}{{\boldsymbol{F}}_{t}}{\boldsymbol{B}} + {\boldsymbol{R}},\\
        {\boldsymbol{K}}_{t-1}&={\boldsymbol{M}}_{t-1}^{-1}{\boldsymbol{B}}^\top {\boldsymbol{F}}_t^\top {\boldsymbol{A}},\\
        l_{t-1}&={\boldsymbol{M}}_{t-1}^{-1}({\boldsymbol{B}}^\top {\boldsymbol{S}}_t+{\boldsymbol{G}_{t-1}}^\top{\boldsymbol{\mu}}_{t-1}),
    \end{split}
\end{equation}
with terminal conditions ${\boldsymbol{F}}_T=\boldsymbol{P}$, ${\boldsymbol{S}}_T=0$, $r_T=0$. 
\end{theorem}
\begin{algorithm}[t]
		\LinesNumbered
		\caption{Primal-dual approach for convex constrained LQR}\label{al:primal-dual}
		\KwIn{initial multiplier ${\boldsymbol{\lambda}}^0\ge 0,{\boldsymbol{\mu}}^0 \ge 0$, a series of step-sizes $\alpha^k$, tolerance $\epsilon$
		}
		\KwOut{multiplier ${\boldsymbol{\lambda}}^k,{\boldsymbol{\mu}}^k$
		\vspace{3pt}	
		}
		\While{$||J({\boldsymbol{u}}^{k+1})-J({\boldsymbol{u}}^k)||>\epsilon$}
		{
		solve ${\boldsymbol{u}}^{k+1}=\mathop {\arg \min }\limits_{\boldsymbol{u}} L({\boldsymbol{u}},{{\boldsymbol{\lambda}} ^k},{\boldsymbol{\mu}}^k)$, s.t. \eqref{model}\\
		update the multiplier ${\boldsymbol{\lambda}}^{k+1}_{i|t}$ and ${\boldsymbol{\mu}}^{k+1}_t$ according to \eqref{gradient}: ${\boldsymbol{\lambda}}^{k+1}_{i|t}=[{\boldsymbol{\lambda}}^{k+1}+\alpha^k\nabla D({\boldsymbol{\lambda}}_{i|t}^k)]_+$, ${\boldsymbol{\mu}}^{k+1}_t=[{\boldsymbol{\mu}}^k+\alpha^k\nabla D({\boldsymbol{\mu}}^k_t)]_+$ for each $t\in \mathcal{T},i\in\mathcal{I}_t$
		}
	\end{algorithm}
\begin{proof}
Since ${\boldsymbol{Q}}_{{\lambda}}-{\boldsymbol{Q}}\succeq0$, and $({\boldsymbol{A}},\sqrt{{\boldsymbol{Q}}})$ is detectable, the pair $({\boldsymbol{A}},\sqrt{{\boldsymbol{Q}}_{{{\lambda}}}})$ is also detectable. 
Under the assumption that $({\boldsymbol{A}}, {\boldsymbol{B}})$ is stabilizable, the optimal cost-to-go is finite, and the optimal control law can stabilize the system. We can then assume that the optimal cost-to-go function $V_t^*(\boldsymbol{x},{\boldsymbol{\lambda}},{\boldsymbol{\mu}})$ takes the quadratic form $\boldsymbol{x}_t^\top {\boldsymbol{F}}_t\boldsymbol{x}_t+{\boldsymbol{S}}_t^\top \boldsymbol{x}_t+r_t$. 
By the HJB equation for a finite time objective, we have:
\begin{equation}\label{HJB}
\begin{split}
	&{\boldsymbol{x}_{t}}^ \top {{\boldsymbol{F}}_{t}}{\boldsymbol{x}_{t}} + {{\boldsymbol{S}}_{t}}^ \top {\boldsymbol{x}_{t}} + {r_{t}} = \\
	&{\mathop {\min }\limits_{\boldsymbol{u}}} [{\boldsymbol{x}_{t-1}}^ \top {{\boldsymbol{F}}_t}{\boldsymbol{x}_{t-1}} + {{\boldsymbol{S}}_{t-1}}^ \top {\boldsymbol{x}_{t-1}} + {r_{t-1}} + {g_{t}}(\boldsymbol{x},\boldsymbol{u},{\boldsymbol{\lambda}} ,{\boldsymbol{\mu}} )].
\end{split}
\end{equation}
Setting the derivative of $\boldsymbol{u}$ over $t$ to zero yields \eqref{law}.
Substituting the optimal control law in \eqref{law} for $\boldsymbol{u}_t$ in \eqref{HJB}, and noticing that the quadratic, linear and constant terms are the same for both sides of the equation, results in \eqref{recursion}.
\end{proof}
Theorem \ref{theorem1} presents the discrete recursive law for the constrained LQR with fixed ${\boldsymbol{\lambda}}$ and ${\boldsymbol{\mu}}$. The method for fixing ${\boldsymbol{\lambda}}$ and ${\boldsymbol{\mu}}$ is shown in Subsection \ref{sub-PDA}.

\begin{proposition}\label{pro1}
 Suppose that all the constraints are inactive within time interval $[k,T]$, then the optimal control law simplifies into $\boldsymbol{u}_t^*(\boldsymbol{x},0,0)=-({\boldsymbol{R}}+{\boldsymbol{B}}^\top {\boldsymbol{F}}_{t+1}{\boldsymbol{B}})^{-1}{\boldsymbol{B}}^\top {\boldsymbol{F}}_{t+1}{\boldsymbol{A}}\boldsymbol{x}_t$, where ${\boldsymbol{F}}_{t+1}$ is the solution of the algebraic Riccati equation $\dot {\boldsymbol{P}}+\boldsymbol{P}{\boldsymbol{A}}+{\boldsymbol{A}}^\top \boldsymbol{P}-\boldsymbol{P}{\boldsymbol{B}}{\boldsymbol{R}}^{-1}{\boldsymbol{B}}^\top {\boldsymbol{P}}-{\boldsymbol{Q}}=0$ at time $t+1$. 
\end{proposition}
\begin{proof}
When all the constraints are inactive, the states $\{\boldsymbol{x}_k,\ldots,\boldsymbol{x}_T\}$ satisfy the inequality constraints strictly. Therefore, from Proposition \ref{pro:KKT}, the corresponding dual variables satisfy $\{{\boldsymbol{\lambda}}_k=0,{\boldsymbol{\mu}}_k=0,\ldots,{\boldsymbol{\lambda}}_T=0,{\boldsymbol{\mu}}_T=0\}$. We then have ${\boldsymbol{Q}}_{{{\lambda}}|t}={\boldsymbol{Q}},\forall t\in[k,T]$ in \eqref{recursion}, which immediately implies that the dynamics of ${\boldsymbol{F}}_t$ are given by the standard algebraic Riccati equation. Hence, ${\boldsymbol{S}}_T^\top=0,{\boldsymbol{S}}_{T-1}^\top={\boldsymbol{C}}_{t-1}^\top{{\boldsymbol{\Lambda}} _{t - 1}} + {\boldsymbol{S}}_t^\top {\boldsymbol{A}}
- ({\boldsymbol{S}}_t^\top {\boldsymbol{B}}+{\boldsymbol{\mu}}_{t-1}{\boldsymbol{Q}}_{t- 1}){{\boldsymbol{M}}^{ - 1}}{{\boldsymbol{B}}^\top }{{\boldsymbol{F}}_t}^\top {\boldsymbol{A}}=0,\ldots,{\boldsymbol{S}}_k^\top=0$. With the recursive law we can see that ${\boldsymbol{S}}_t=0,\forall t=k,\ldots,T$.
\end{proof}
Proposition \ref{pro1} gives a theoretical illustration of what happens when the state enters the \textit{invariant set}. In this case, the residual problem can be solved using unconstrained LQR. This is the backbone of solving the infinite horizon-constrained LQR. We then immediately prove that for any ${\boldsymbol{\lambda}},{\boldsymbol{\mu}}\ge 0$, the optimal law \eqref{law} stabilizes the system.
\begin{theorem}
For the given fixed dual variables ${\boldsymbol{\lambda}},{\boldsymbol{\mu}}$, the control law $\boldsymbol{u}_t^*(\boldsymbol{x}_t,{\boldsymbol{\lambda}},{\boldsymbol{\mu}})$ stabilizes the system.
\end{theorem}
\begin{proof}
Since ${\boldsymbol{Q}}_{{{\lambda}}}\succeq {\boldsymbol{Q}}$ and $({\boldsymbol{A}},\sqrt{{\boldsymbol{Q}}})$ is detectable, the pair $({\boldsymbol{A}},\sqrt{{\boldsymbol{Q}}_{{{\lambda}}}})$ is also detectable. The feedback quantity $k_t$ of the control law $\boldsymbol{u}_t^*(\boldsymbol{x},{\boldsymbol{\lambda}})$ is $-({\boldsymbol{R}}+{\boldsymbol{B}}^\top {\boldsymbol{F}}_{t+1}{\boldsymbol{B}})^{-1}{\boldsymbol{B}}^\top {\boldsymbol{F}}_{t+1}{\boldsymbol{A}}$. This implies that the spectral radius $\rho({\boldsymbol{A}}+{\boldsymbol{B}}{\boldsymbol{K}})<0$ proves the asymptotic stability of the closed-loop system.
\end{proof}

\section{Simulation \& Experiment}\label{sec:simulation}

In this section, we verify the proposed method through the following:
\romannumeral1) the effectiveness of the algorithm in terms of the trajectory's feasibility and stability of the equilibrium;
\romannumeral2) the computation time compared to that of nonlinear solvers, i.e., interior-point (Ipopt), SQP, SQP-legacy (SQP-L), active-set (act-set), and iSCA;
the computation time compared to that of motion planners, i.e., RRT*, control-based RRT (C-RRT), PRM, A* and Hybrid A* (H-A*).
\romannumeral3) feasibility comparison with the above solvers and energy cost comparison with the above planners.

\begin{figure}[t]
	\centering
	\includegraphics[scale=0.5]{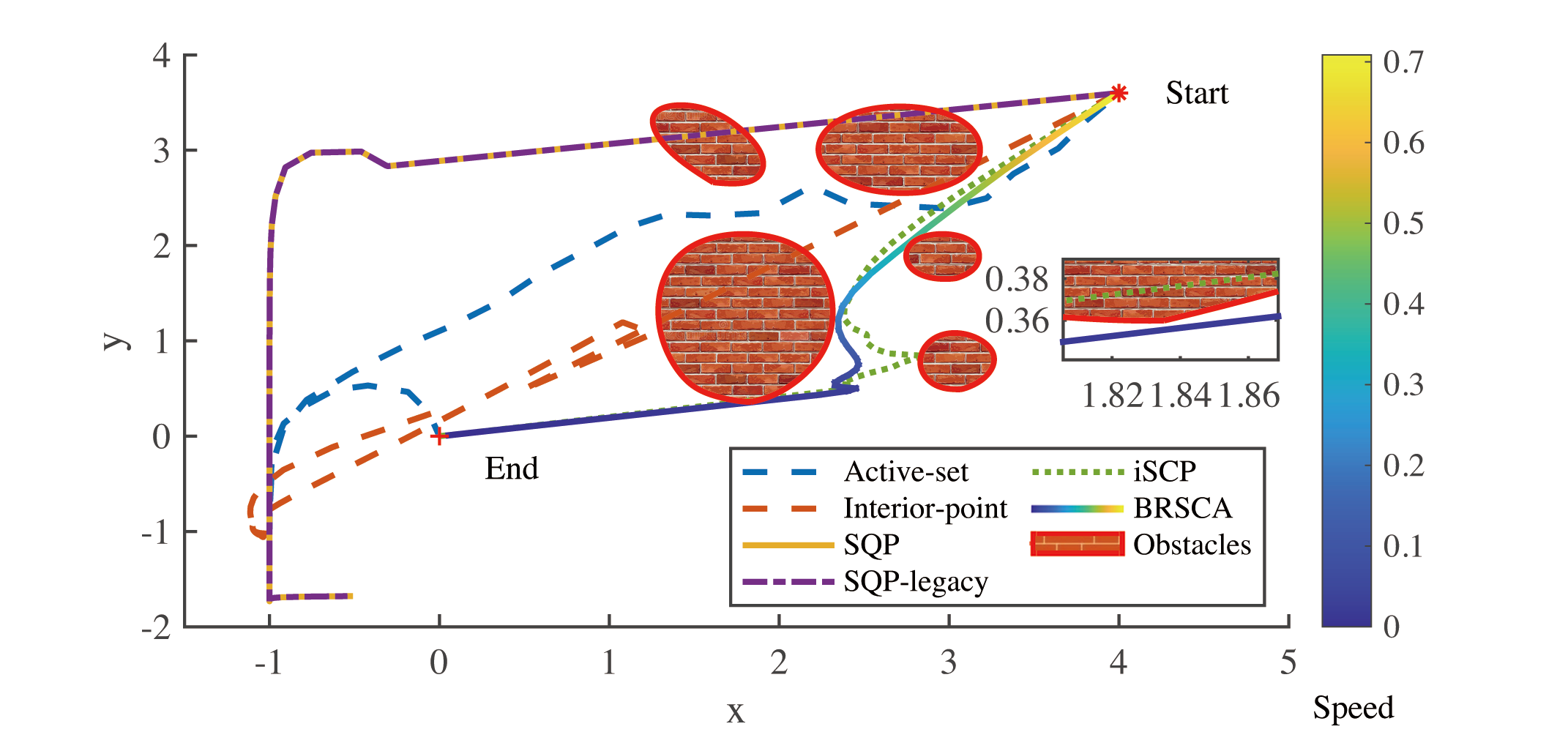}
	\caption{Comparison between BRSCA and nonlinear solvers, iSCA with five obstacles in numerical simulation. BRSCA and Active-set method achieve the collision-free task.}
	\label{fig:5obs}
\end{figure}
\begin{figure}[t]
	\centering
	\includegraphics[scale=0.5]{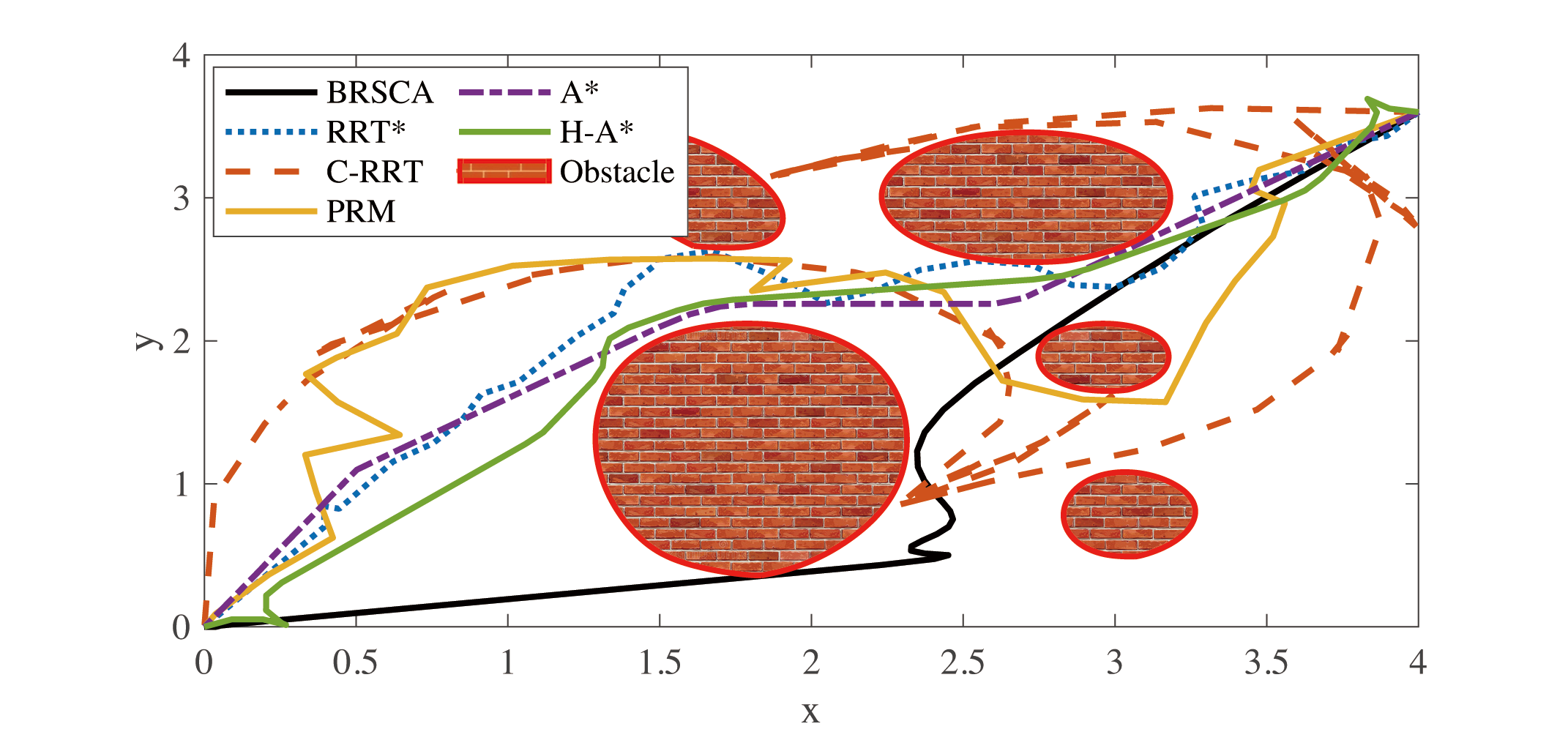}
	\caption{Comparsion between BRSCA and planners with five obstacles in numerical simulation. BRSCA has the lowest energy cost.}
	\label{fig:5obs-planners}
	\vspace{-10pt}
\end{figure}

We consider three sets of experiments.
In the first and second sets, we consider scenarios with 5 and 15 random irregular-shaped semi-convex obstacles along with a 100-length planning horizon for numerical verification. In the third set, we test our method with an Omnidirectional robot on the testbed\cite{ding2021robopheus} to verify the effectiveness in practice.
The testbed involves seven irregular-shaped obstacles.
The robot's goal is to reach the endpoint stably without colliding with the obstacle and minimize the energy cost.
The solvers are used to compute the control input directly.
The planners are used to generate the reference trajectories. Then a P-controller is used to compute the control input.
%


All numerical experiments are performed in MATLAB on a computer with an Intel(R) Core(TM) i9-9980XE CPU, 3.00GHz processor and 64GB RAM. The hardware implementation on the testbed is performed with the Robopheus reality testbed\cite{ding2021robopheus}.

\subsection{Numerical Simulation}
In the first scenario, the workspace is a 4 $\times$ 4 square with five irregular-shaped obstacles randomly placed inside. The obstacle coverage rate is 44.3\% at the central area (from [1.15, 0.3] to [3.36, 3.6]). The tolerance $\epsilon$ in Algorithm \ref{al:primal-dual} is $\epsilon = 0.7$. The length of the planning horizon is set to 100. The start point is $[4, 3.6]$, and the desired endpoint is $[0, 0]$. The control input is bounded by a box constraint $[-0.7,0.7]$ on both $x$ and $y$ directions. The trajectory and speed calculated by the proposed method are shown in Fig.\ref{fig:5obs} and Fig.\ref{fig:5obs-planners}. 

\begin{figure}[t]
	\centering
	\includegraphics[scale=0.5]{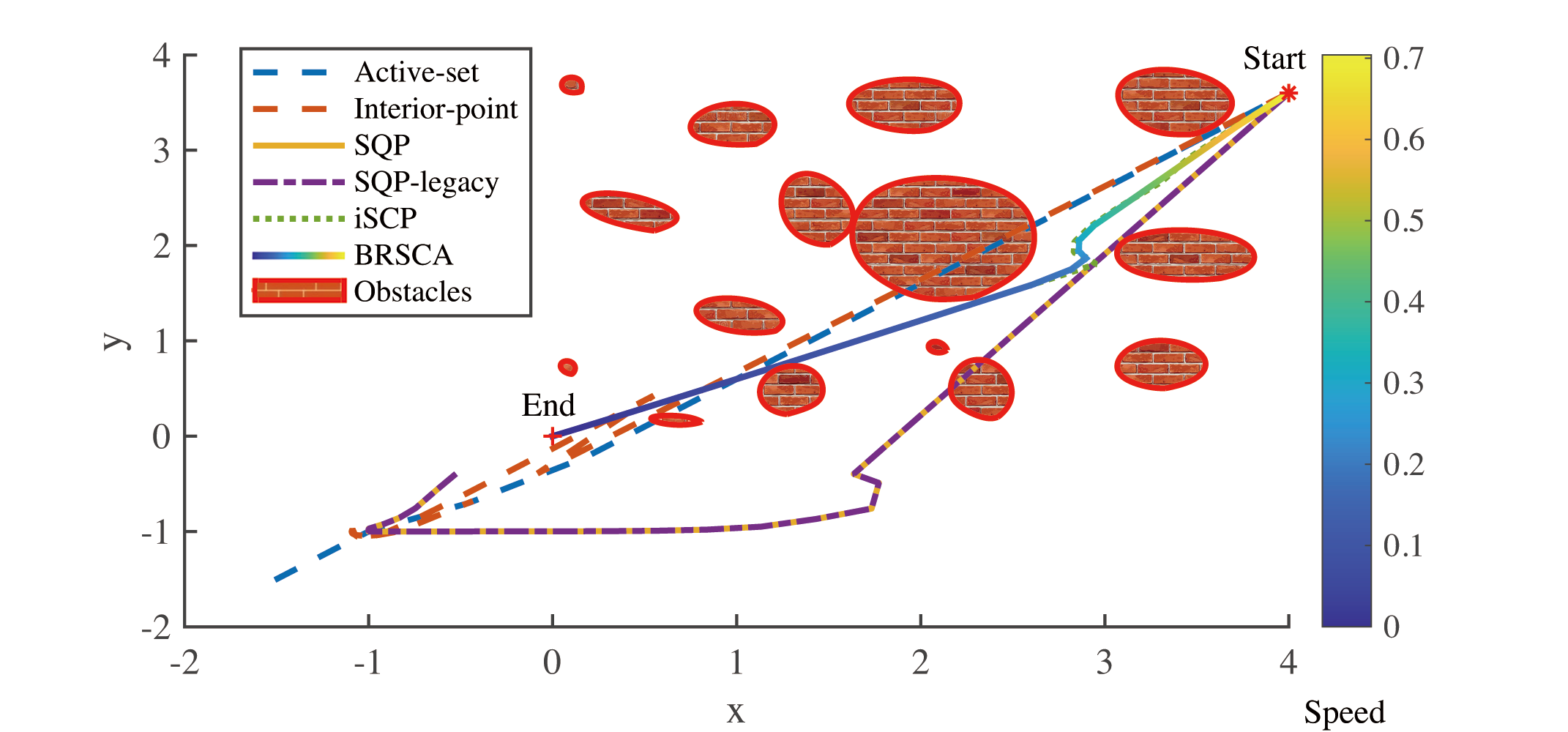}
	\caption{Comparsion between BRSCA and nonlinear solvers, iSCA with 15 obstacles in numerical simulation. BRSCA and iSCA method achieve the collision-free task. }
	\label{fig:15obs}
\end{figure}
\begin{figure}[t]
	\centering
	\includegraphics[scale=0.5]{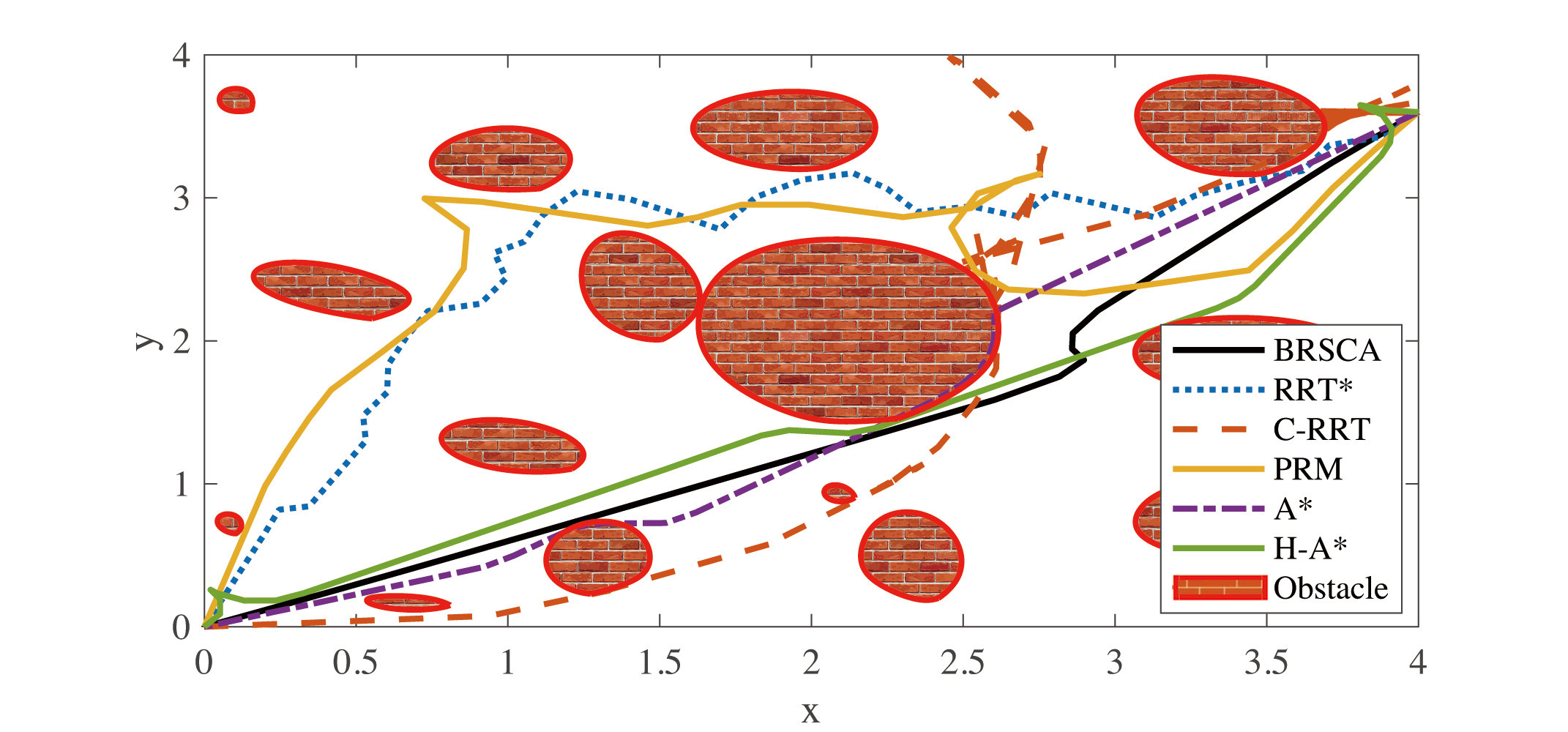}
	\caption{Comparsion between BRSCA and planners with 15 obstacles in numerical simulation. BRSCA has the lowest energy cost.}
	\label{fig:15obs-planners}
	\vspace{-10pt}
\end{figure}

In Fig.\ref{fig:5obs}, the active-set solver is the only method that produced a collision-free trajectory among the nonlinear solvers used. Our method also results in a collision-free trajectory, with a much smaller cost (96.02 compared to 283.96 of the trajectory returned by the active-set solver). The iSCA is infeasible at some points. Details of infeasibility are shown in the magnification, where the dot, red, and coloured line denote the trajectory of iSCA, the boundary of the obstacle and the trajectory of BRSCA, respectively. As discussed, infeasibility is caused by possible linearisation about included infeasible points.
In Fig.\ref{fig:5obs-planners}, all the planners produced collision-free trajectories. 
The RRT* has the shortest computing time with 0.068s (compared to 0.74s of the BRSCA).
However, the costs of the planners are much more than the proposed BRSCA.
BRSCA reduces the energy cost by at least 449.78\% compared to other planning methods.
The detailed comparisons of computing time, cost and collision-free rate of the BRSCA and the planners are shown in Tab.\ref{tb:time_planner}.
The collision-free rate is defined as the number of the collision-free trajectory against the total number of the trajectory.

In the second scenario, all the settings are the same as those of the first scenario, except that the number of obstacles is 15.
Two obstacles in the middle of the scenario combine a nonconvex obstacle and consist of a \textit{semi-convex} set.
As shown in Fig.\ref{fig:15obs}, BRSCA provided a feasible trajectory and stable control for obstacle avoidance when encountering more constraints. 
The iSCA is the only other method that resulted in a collision-free trajectory. The cost value of iSCA is 98.4897, which is slightly bigger than our 96.7878.
The other methods did not perform well when 1700 constraints were included (the trajectory is infeasible). 
In Fig.\ref{fig:15obs-planners}, BRSCA, RRT*, PRM and H-A* produced collision-free trajectories. 
BRSCA reduces the cost by at least 464.23\% compared to other planning methods.
The detailed comparisons of computing time, cost and collision-free rate of the BRSCA and the planners are shown in Tab.\ref{tb:time_planner}.

\subsection{Hardware Implementation}

\begin{figure}[t]
    \centering
    \includegraphics[scale=0.25]{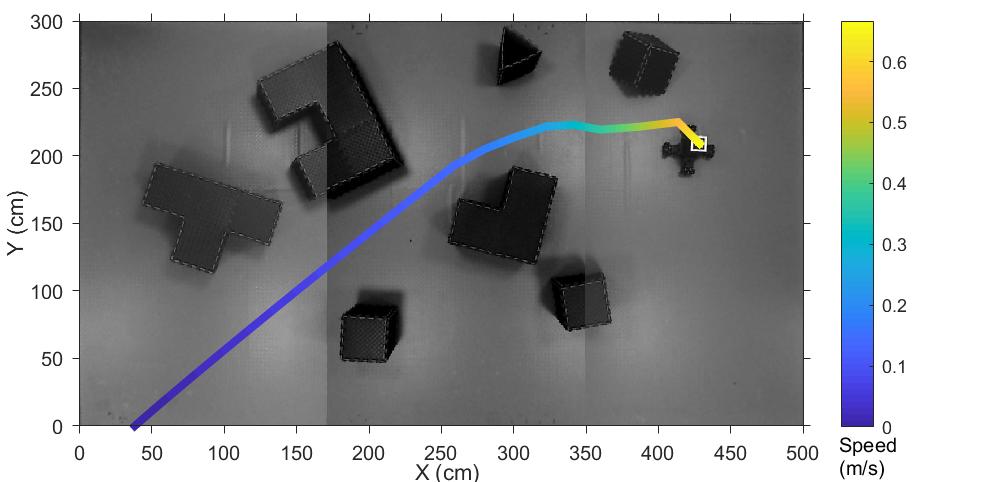}
    \caption{Experiment with seven obstacles}
    \label{fig:realexp}
    \vspace{-10pt}
\end{figure}

We validate BRSCA on a hardware testbed, which is a 5m $\times$ 3m rectangular space with seven irregular obstacles randomly placed inside. The obstacle coverage rate is about 9.3\%. The tolerance in Algorithm \ref{al:primal-dual} is set to $\epsilon = 0.1$, and the length of the planning horizon in this scenario is 200.
The start point is randomly set at $[429, 207]$, and the endpoint is $[38, 0]$ but not the origin in this case to avoid edge distortion of the cameras on the boundary. The control frequency is 25Hz. Fig.\ref{fig:realexp} depicts the collision-free trajectory and the velocity variations. The diminished velocity demonstrates stability around the endpoint.
The video link: \url{https://www.youtube.com/watch?v=c2bw2O7EfDA}.

\subsection{Computation Time}
Table \ref{tb:time1} compares the computation time between BRSCA and nonlinear solvers, iSCA. It can be seen that BRSCA can solve the problem with multiple obstacles within around 1 second. Besides, the computation speed grows linearly with the number of obstacles empirically. The computation speed grows moderately with the number of obstacles or even decreases for other algorithms because they fail to find feasible solutions, e.g. SQP with 15 obstacles. The performance of BRSCA is illustrated better, even though the comparison is somewhat unfair to ours. In summary, BRSCA has a higher probability of finding feasible solutions and reduces the computation time by at least 17.4\%. 

\begin{table}[h]
	\caption{Computing time, cost and collision-free rate of different planners}
	\begin{tabular}{ccccccc}
		\toprule[1.5pt]                                          & BRSCA          & RRT*           & C-RRT  & PRM    & A*     & H-A* \\
		\midrule[0.75pt]
		\begin{tabular}[c]{@{}c@{}}Time\\ (5 Obs)\end{tabular}        & 0.74           & \textbf{0.068} & 1.40   & 0.14   & 1.97   & 1103.97   \\
		\begin{tabular}[c]{@{}c@{}}Time\\ (15 Obs)\end{tabular}       & 1.36           & \textbf{0.061} & 0.42   & 0.14   & 1.93   & 157.79    \\
		\begin{tabular}[c]{@{}c@{}}Cost \\ (5 Obs)\end{tabular}       & \textbf{96.02} & 431.88         & 1739.0 & 465.95 & 470.96 & 666.60    \\
		\begin{tabular}[c]{@{}c@{}}Cost \\ (15 Obs)\end{tabular}      & \textbf{96.78} & 478.78         & 2011.3 & 497.93 & 449.28 & 678.74    \\
		\begin{tabular}[c]{@{}c@{}}Collision\\ free rate\end{tabular} & 100\%          & 100\%          & 50\%   & 100\%  & 50\%    & 100\%  \\
		\bottomrule[1.5pt]
	\end{tabular}
	\label{tb:time_planner}
\end{table}

\begin{table}[h]
\caption{Computing time and collision-free rate of different solvers}
\begin{center}
\begin{threeparttable}
\begin{tabular}{ccccccc}
\toprule[1.5pt]
                                                            & BRSCA          & iSCA & Ipopt & SQP    & SQP-L  & Act-set \\ \midrule[0.75pt]
5 Obs                                                             & \textbf{0.74}  & 1.30 & 9.88  & 9.17   & 10.67  & 163.89  \\
7 Obs                                                             & \textbf{0.98}  & 1.24 & 9.68  & 31.38  & 38.21  & 70.28   \\
9 Obs                                                             & \textbf{1.04}  & 1.26 & 10.47 & 91.24  & 124.13 & 70.58   \\
12 Obs                                                            & \textbf{1.23}  & 1.31 & 11.83 & 122.22 & 174.16 & 76.67   \\
15 Obs                                                            & \textbf{1.36}  & 1.37 & 13.49 & 27.91  & 38.48  & 220.48  \\ \midrule[0.75pt]
Time\tnote{*}                                                           & \textbf{1.00}  & 1.21 & 10.35 & 52.71  & 72.10  & 112.54  \\
\begin{tabular}[c]{@{}l@{}}Collision \\ free rate\end{tabular} & \textbf{100\%} & 60\% & 60\%  & 0\%    & 0\%    & 0\%     \\ \bottomrule[1.5pt]
\end{tabular}

\begin{tablenotes}
        \item[*] The average time compared to BRSCA.
      \end{tablenotes}    
      \end{threeparttable}
\end{center}

\label{tb:time1}
\end{table}

\begin{table}[h]
\caption{Computation time with different tolerance settings and numbers of obstacles}
\begin{center}
\begin{tabular}{cccccc}
\toprule[1.5pt]
          $\epsilon$ & 5 obs         & 7 obs       & 9 obs         & 12 obs       & 15 obs      \\
          \midrule[0.75pt]
0.0003 & 58.78 & 70.15 & 80.71 & 96.78 & 114.31 \\
0.03   & 31.89 & 38.58 & 44.86 & 54.39 & 64.66 \\
0.7    & 0.74    & 0.98 & 1.04  & 1.23  & 1.36 \\
\bottomrule[1.5pt]
\end{tabular}
\end{center}
\label{tb:time}
\end{table}
Table \ref{tb:time} shows the computation time of BRSCA with different tolerance settings and numbers of obstacles. With a modest tolerance such as 0.7, BRSCA is real-time implementable. We note that an appropriate stepsize selection in Algorithm \ref{al:primal-dual} can realize further acceleration. Our future work will investigate selections of stepsize and an accelerated primal-dual approach.
\section{Conclusion}\label{sec:conclusion}
This paper develops the BRSCA algorithm for motion planning using linear quadratic regulator formulation. We demonstrate the relative merits of our approach with incremental SCA, which uses a similar principle. We also explore the special structure of the formulated constrained LQR problem by transforming it into a convex QCQP. We prove that the dual Lagrangian problem can be regarded as an unconstrained LQR for every dual variable. The unconstrained LQR is proved to have stability guarantees for every dual variable, thereby proving stability during primal-dual iterations. For every dual variable, we use backward recursion to give closed-form solutions for the optimal cost-to-go and feedback control law. The proposed algorithm is validated on a hardware testbed and through numerical simulations. 

Despite the advantages of the proposed method demonstrated
above, it also has some limitations that need to be
studied in future research.
1) Nonlinear system. When facing a nonlinear dynamic system, BRSCA can be used after linearisation about equilibrium points.
A sub-optimal control can be obtained with a state-dependent Riccati equation approach.
2) Moving obstacles. Our problem formulation \eqref{SCLQR} does not rely on stable obstacles. As long as $h_i(\boldsymbol{x}_t)$ is known, BRSCA can solve the problem.
However, it is difficult to know the dynamic of the moving obstacle in practice.
There are two possible ways to solve this.
i) The prediction of the obstacles can be used in BRSCA.
ii) A low-level control filter can be designed to correct the reference trajectory made by BRSCA and avoid collision based on sampling with onboard sensors.




\bibliographystyle{ieeetr}
\bibliography{article.bib}
\end{document}